\documentstyle[12pt] {article}
\newcommand{\beq}{\begin{equation}}
\newcommand{\eeq}{\end{equation}}
\newcommand{\beqa}{\begin{eqnarray}}
\newcommand{\eeqa}{\end{eqnarray}}
\newcommand{\ba}{\begin{array}}
\newcommand{\ea}{\end{array}}

\begin{document}

\begin{flushright}
Preprint DFPD/97/TH/15
\end{flushright}

\begin{center}
{\large \bf On the Torus Quantization of Two Anyons \\
with Coulomb Interaction in a Magnetic Field} 
\end{center}

\vskip 1. truecm

\begin{center}
{\bf Luca Salasnich}\footnote{Electronic address: 
salasnich@math.unipd.it}
\vskip 0.5 truecm
Dipartimento di Matematica Pura ed Applicata \\
Universit\`a di Padova, Via Belzoni 7, I 35131 Padova, Italy \\
and\\
Istituto Nazionale di Fisica Nucleare, Sezione di Padova,\\
Via Marzolo 8, I 35131 Padova, Italy
\end{center}

\vskip 1. truecm

\begin{center}
{\bf Abstract}
\end{center}

\vskip 0.5 truecm
\par
We study two anyons with Coulomb interaction in a uniform 
magnetic field $B$. By using the torus quantization 
we obtain the modified Landau and Zeeman formulas for the two anyons. 
Then we derive a simple algebraic equation for the full spectral 
problem up to the second order in $B$. 

\vskip 0.5 truecm
\begin{center}
To be published in {\it Modern Physics Letters} B
\end{center} 

\newpage

\par
In 1977 Leinaas and Myrheim$^{1)}$ suggested 
the possible existence of arbitrary statistics 
between the Bose and the Fermi case in 
space dimensions less than three.  
A model of particles with
arbitrary, or fractional, statistics was introduced 
by Wilczek$^{2)}$, who called such objects as {\it anyons}. 
Anyons are thought to be good 
candidates for the explanation of such important
collective phenomena in condensed matter physics as the
fractional quantum Hall effect and high temperature
superconductivity (for recent reviews see Ref. 3--5). 
\par
In this paper we study two anyons with Coulomb interaction in a uniform 
magnetic field. The eigenvalue problem for this system 
was solved numerically by 
Myrheim, Halvorsen and Vercin$^{6)}$ by using a finite difference 
method to discretize the Schr\"odinger equation. 
Here, instead, we obtain some analytical formulas of the 
energy spectrum by using the torus quantization$^{7),8)}$. 
\par
The Lagrangian of two anyons with Coulomb 
interaction in two dimensions is given by
\beq
L_0={m\over 2}({\dot {\bf r}_1}^2+{\dot {\bf r}_2}^2) 
+ {e^2 \over ||{\bf r}_2 - {\bf r}_1||} 
+ \alpha \hbar {\dot \theta} 
\eeq
where ${\bf r}_1=(x_1,y_1)$, ${\bf r}_2=(x_2,y_2)$ and 
$\alpha$ is the statistical parameter: Bose statistics is recover for 
$\alpha =0$ and Fermi statistics for $\alpha =1$. The azimuthal 
angle is defined by $\theta = \arctan ({y_2 - y_1 \over x_2 - x_1})$. 
\par
The separation of the center of mass motion can be achieved through 
the change of variables ${\bf r} = {\bf r}_2 - {\bf r}_1$ and 
${\bf r}_G = {\bf r}_2 + {\bf r}_1$. In this way we obtain 
\beq
L_0= {\mu \over 2} {\dot {\bf r}_G}^2 
+ {\mu \over 2} {\dot {\bf r}}^2 + {e^2\over r} 
+ \alpha \hbar {\dot \theta} \; ,
\eeq
where $\mu = m/2$ is the reduced mass and $r=||{\bf r}||$. 
Neglecting the motion of the center of mass ${\bf r}_G$, 
the Lagrangian can be written in polar coordinates 
${\bf r}=(r \cos{\theta}, r\sin{\theta})$ as 
\beq
L_0={\mu \over 2} ({\dot r}^2 + r^2 {\dot \theta}^2 ) + {e^2\over r} 
+ \alpha \hbar {\dot \theta} \; .
\eeq
\par
Let us introduce a uniform magnetic field ${\bf B}$ along the 
$z$--axis. Choosing the symmetric gauge, 
the Lagrangian of interaction is given by 
\beq
L_I = -{e B\over 2 c} r^2 {\dot \theta} \; ,
\eeq
so that the total lagrangian reads 
\beq
L = L_0 + L_I = {\mu \over 2} ({\dot r}^2 + r^2 {\dot \theta}^2 ) 
+ {e^2\over r} + (\alpha \hbar - 
{eB\over 2c}r^2 ){\dot \theta} \; .
\eeq
Introducing the generalized momenta 
\beq
p_r = {\partial L\over \partial {\dot r} } = \mu {\dot r} \; ,
\eeq
\beq
p_{\theta}={\partial L\over \partial {\dot \theta} } =
\mu r^2 {\dot \theta} + (\alpha \hbar -{eB\over 2c}r^2 ) \; ,
\eeq
the Hamiltonian $H=p_r {\dot r} + p_{\theta} {\dot \theta} - L$ 
can be written as 
\beq
H = {p_r^2\over 2 \mu} + {(p_{\theta}-\alpha \hbar )^2\over 2 \mu r^2}
- {e^2\over r} + {e^2 B^2 \over 8 \mu c^2}r^2 
+ {eB\over 2 \mu c} (p_{\theta} - \alpha \hbar ) = E \; .
\eeq
The angular momentum $p_{\theta}$ is a constant of motion 
because the Hamiltonian is independent of $\theta$. 
It is readily verified that $H$ and $p_{\theta}$ are in involution, 
so this two--degrees of freedom system is integrable. 
We observe that ${eB\over 2 \mu c} (p_{\theta} - \alpha \hbar )$  
is constant and we study the reduced Hamiltonian 
\beq
{\bar H} = {p_r^2\over 2 \mu} + W(r) = {\bar E} ,
\eeq
where 
\beq
W(r)= {(p_{\theta}-\alpha \hbar )^2\over 2 \mu r^2}
- {e^2\over r} + {e^2 B^2 \over 8 \mu c^2}r^2        \; . 
\eeq
Also the Hamiltonian ${\bar H}$ is integrable and each classical 
trajectory is confined on a 2--dimensional torus. 
The action variables on the torus are given by
\beq
I_{\theta}={1\over 2\pi}\oint p_{\theta} d\theta = p_{\theta} \; , 
\eeq
\beq
I_r = {1\over 2\pi} \oint \sqrt{2\mu ({\bar E} - W(r))} dr \; ,
\eeq
and by inverting the last relation 
we have ${\bar E}={\bar H}(I_r, I_{\theta})$. 
\par
The energy spectrum of the system can be obtained by using 
the {\it torus quantization}. 
The torus quantization goes back to the early days of quantum mechanics 
and was developed by Bohr and Sommerfeld for 
separable systems, it was then generalized for integrable 
(but not necessarily separable) systems by Einstein$^{7)}$. 
In fact, Einstein's result was corrected for the phase changes 
due to caustics by Maslov$^{8)}$. 
The torus quantization 
is just the first term of a certain $\hbar$-expansion 
of the Schr\"odinger equation$^{8),9)}$ but it is 
of extreme importance since in many cases this is the only 
approximation known in the form of an explicit formula$^{10),11)}$. 
For our system the torus quantization reads 
\beq
I_{\theta} = n_{\theta} \hbar \; , 
\eeq
\beq
I_r = (n_r +{1\over 2}) \hbar \; ,
\eeq 
where the radial quantum number $n_r$ is a non--negative integer, 
whilst the angular quantum number $n_{\theta}$ can be also negative. 
\par
In this way the quantized spectrum is given by
\beq
E(n_r, n_{\theta})= {\bar E}(n_r, n_{\theta}) + 
{eB\hbar \over 2\mu c}(n_{\theta} - \alpha ) \; ,
\eeq
where ${\bar E}={\bar E}(n_r ,n_{\theta})$ is obtained by solving the equation 
\beq
{1\over 2\pi}\oint \sqrt{ - F + {2 G\over r} - {M \over r^2} - N r^2 } dr
= (n_r+{1\over 2})\hbar  \; ,
\eeq
with $F=2\mu (-{\bar E})$, $G=\mu e^2$, $M=(n_{\theta} - \alpha )^2 \hbar^2$ 
and $N=e^2B^2/(4c^2)$. 
\par
This integral equation can be evaluated analytically 
in some limit cases. In particular, if $B$ is very strong then 
the Coulomb interaction is negligible ($G=0$). In this case, 
if $r$ is represented in the complex plane, the integrand can be pictured 
on a Riemann surface of two sheets with branch points at the 
roots $r_1$ and $r_2$ ($r_1<r_2$) of the radicand$^{12)}$. 
The path of integration 
encloses the line joining the two roots but, with the inverse 
direction of rotation, the path encloses the poles of the 
integrand: $r=0$ and $r=\infty$. It follows from the 
{\it theorem of residues} that the integral 
is equal to the negative sum of the residues $Res_0$ and 
$Res_{\infty}$ of the integrand in these poles 
\beq
\oint \sqrt{ - F - {M \over r^2} - N r^2 } dr 
= - 2 \pi i (Res_0 + Res_{\infty} ) 
= - 2\pi (\sqrt{M} + {F\over 2\sqrt{N}})       \; ,
\eeq
where $Res_0$ is the coefficient of $1/r$ 
in the Laurent expansion of the integrand 
in the neighbourhood of the pole $r=0$. Instead $Res_{\infty}$ is 
calculated as the $Res_0$ of the integrand arising from the 
substitution $z=1/r$. 
In this way we obtain the modified Landau formula for the two anyons 
\beq
E_L(n_r, n_{\theta}) = {eB\hbar \over 2\mu c} 
(n_r + {1\over 2} + |n_{\theta} - \alpha | + (n_{\theta} - \alpha ))  \; . 
\eeq
\par
Now we consider the case of a week magnetic field. 
In the simplest case $B^2$ is negligible ($N=0$) and 
we can apply again the theorem of residues
\beq 
\oint \sqrt{ - F + {2 G\over r} - {M \over r^2} } dr 
= - 2\pi (\sqrt{M} - {G\over \sqrt{F}} )  \; ,
\eeq
from which we obtain the modified Zeeman formula for the anyons 
\beq
E_Z(n_r, n_{\theta}) = { - \mu e^4 \over 2 
\hbar^2 [(n_r + {1\over 2})+|n_{\theta} - \alpha |]^2} 
+ {eB\hbar \over 2\mu c}(n_{\theta} - \alpha )        \; .
\eeq
\par
Let us study the case $N\neq 0$ where $N r^2$ represents a 
correction term. We expand the square root of Eq. (16) in powers of $N$. 
At first order in $N$ (second order in $B$) we get 
\beq
\oint \sqrt{ - F + {2 G\over r} - {M \over r^2} - N r^2 } dr =
\oint \sqrt{ - F + {2 G\over r} - {M \over r^2} } dr 
- {N\over 2} \oint {r^3 \over \sqrt{-Fr^2 + 2G r - M} } dr \; .
\eeq 
The first integral has been obtained previously and the second one gives
\beq 
\oint {r^3 \over \sqrt{-Fr^2 + 2G r - M} } dr =
- 2\pi (3 M - {5G^2\over F}) {G\over 2 F^2 \sqrt{F}}  \; .
\eeq
The quantization formula, up to the second order in 
$B$, is given by Eq. (15) where ${\bar E}$ is obtained 
by solving the following equation 
\beq
{\bar E}^7 = 
{- \mu e^4 \over 2 \hbar^2 [(n_r + {1\over 2}) + |n_{\theta} - \alpha |]^2 }
\Big( {\bar E}^3 + 
{3 e^2 B^2 (n_{\theta}-\alpha )^2 \hbar^2 \over 64 \mu^2 c^2} {\bar E} +
{5 e^6 B^2 \over 128 \mu c^2} \Big)^2  \; .
\eeq
This algebraic equation, which can be easily solved numerically 
by using the Newton--Raphson method, 
gives the first correction to the modified Zeeman formula. 
To get higher order corrections it is sufficient to calculate 
other terms in the expansion of the square root of Eq. (16). 
\par
In summary, we have obtained some analytical formulas for the 
energy spectrum of two anyons with Coulomb 
interaction in a uniform magnetic field $B$. 
The system has been quantized by using the torus quantization 
and the integrals which derive from it have been calculated 
in the complex plane by using the theorem of residues. 
In the limit of strong magnetic field we have derived a 
modified Landau formula. Instead, for a weak magnetic field 
we have deduced a modified Zeeman formula and then its correction 
up to the second order in $B$. 

\vskip 0.5 truecm

\begin{center}
*****
\end{center}
\par
The author is greatly indebted to T.A. Minelli, M. Robnik and 
F. Sattin for many enlightening discussions. 

\newpage
\section*{References}
\noindent

1. J. M. Leinaas and J. Myrheim, {\it Nuovo Cimento}
{\bf B 37}, 1 (1977). 

2. F. Wilczek, {\it Phys. Rev. Lett.} {\bf 48}, 1144
(1982). 

3. S.M. Girvin and R. Prange, {\it The Quantum Hall 
Effect} (Springer, New York, 1990). 

4. F. Wilczek, {\it Fractional Statistics and Anyon 
Superconductivity} (World Scientific, Singapore, 1991). 

5. S. Forte, {\it Rev. of Mod. Phys.} {\bf 64}, 193 (1992). 

6. J. Myrheim, E. Halvorsen and A. Vercin, {\it Phys. Lett.} {\bf B 278}, 
171 (1992).

7. A. Einstein, {\it Verh. Dtsch. Phys. Ges.} {\bf 19}, 82 (1917).

8. V.P. Maslov and M.V. Fedoriuk, {\it Semi-Classical Approximations in 
Quantum Mechanics} (Reidel Publishing Company, Boston, 1981). 

9. M. Robnik and L. Salasnich, {\it J. Phys.} A {\bf 30}, 1711 (1997).

10. V.R. Manfredi and L. Salasnich, {\it Int. J. Mod. Phys.} {\bf B 9}, 
3219 (1995).

11. L. Salasnich, {\it Phys. Rev.} {\bf D 52}, 6189 (1995).

12. A. Voros, {\it Ann. Inst. H. Poincar\'e} {\bf A 39}, 211 (1983). 

\end{document}